\documentclass[letterpaper]{article} 
\usepackage{aaai24}  
\usepackage{times}  
\usepackage{helvet}  
\usepackage{courier}  
\usepackage[hyphens]{url}  
\usepackage{graphicx} 
\usepackage{natbib}  
\usepackage{caption} 
\usepackage{algorithm}
\usepackage{algorithmic}

\usepackage{amsmath} 

\usepackage{newfloat}
\usepackage{listings}
\DeclareCaptionStyle{ruled}{labelfont=normalfont,labelsep=colon,strut=off} 
\floatstyle{ruled}
\newfloat{listing}{tb}{lst}{}
\floatname{listing}{Listing}
\newcommand{\country}[1]{\textit{#1}}

\title{Temporal Graph Networks for Graph Anomaly Detection in Financial Networks}

\author{
    Yejin Kim\textsuperscript{1}\thanks{These authors contributed equally.},
    Youngbin Lee\textsuperscript{1}\footnotemark[1],
    Minyoung Choe\textsuperscript{2},
    Sungju Oh\textsuperscript{3},
    Yongjae Lee\textsuperscript{1}\thanks{Corresponding author.}
}

\affiliations {

    \textsuperscript{\rm 1}Ulsan National Institute of Science and Technology, \country{South Korea}\\
    \textsuperscript{\rm 2}Korea Advanced Institute of Science and Technology, \country{South Korea}\\
    \textsuperscript{\rm 3}Sogang University, \country{South Korea}\\

    \{kimyejin99, young, yongjaelee\}@unist.ac.kr, minyoung.choe@kaist.ac.kr, osjsj0210@sogang.ac.kr

}

\begin{document}

\maketitle

\begin{abstract}
This paper explores the utilization of Temporal Graph Networks (TGN) for financial anomaly detection, a pressing need in the era of fintech and digitized financial transactions. We present a comprehensive framework that leverages TGN, capable of capturing dynamic changes in edges within financial networks, for fraud detection. Our study compares TGN's performance against static Graph Neural Network (GNN) baselines, as well as cutting-edge hypergraph neural network baselines using DGraph dataset for a realistic financial context. Our results demonstrate that TGN significantly outperforms other models in terms of AUC metrics. 
This superior performance underlines TGN's potential as an effective tool for detecting financial fraud, showcasing its ability to adapt to the dynamic and complex nature of modern financial systems. 
We also experimented with various graph embedding modules within the TGN framework and compared the effectiveness of each module. In conclusion, we demonstrated that, even with variations within TGN, it is possible to achieve good performance in the anomaly detection task.
\end{abstract}

\section{Introduction}

Detecting financial fraud is crucial in the evolving landscape of the financial sector, especially as transactions become more digitized and complex, particularly with the rise of fintech. The financial industry's high reliance on trust makes fraud detection vital for upholding system integrity and preventing significant financial losses to individuals, businesses, and financial institutions. Traditional methods like manual review, rule-based systems, and auditing prove insufficient in addressing the diverse and advanced methods employed in financial fraud \cite{west2016intelligent}. Consequently, there's a notable shift towards leveraging AI and big data for fraud detection. \cite{ngai2011application}

The landscape of financial transactions has been profoundly transformed by the rise of fintech, with a significant majority now being conducted via mobile devices \cite{kim2016adoption}. This shift has not only increased the volume and velocity of financial activities but also introduced complex challenges in real-time fraud detection. In this context, real-time monitoring systems, particularly those leveraging interaction graph databases, have become pivotal in modeling the intricate relationships between entities such as users, accounts, and transactions \cite{altman2023realistic}.

\begin{figure}[h]
    \centering
    \includegraphics[width=\linewidth]{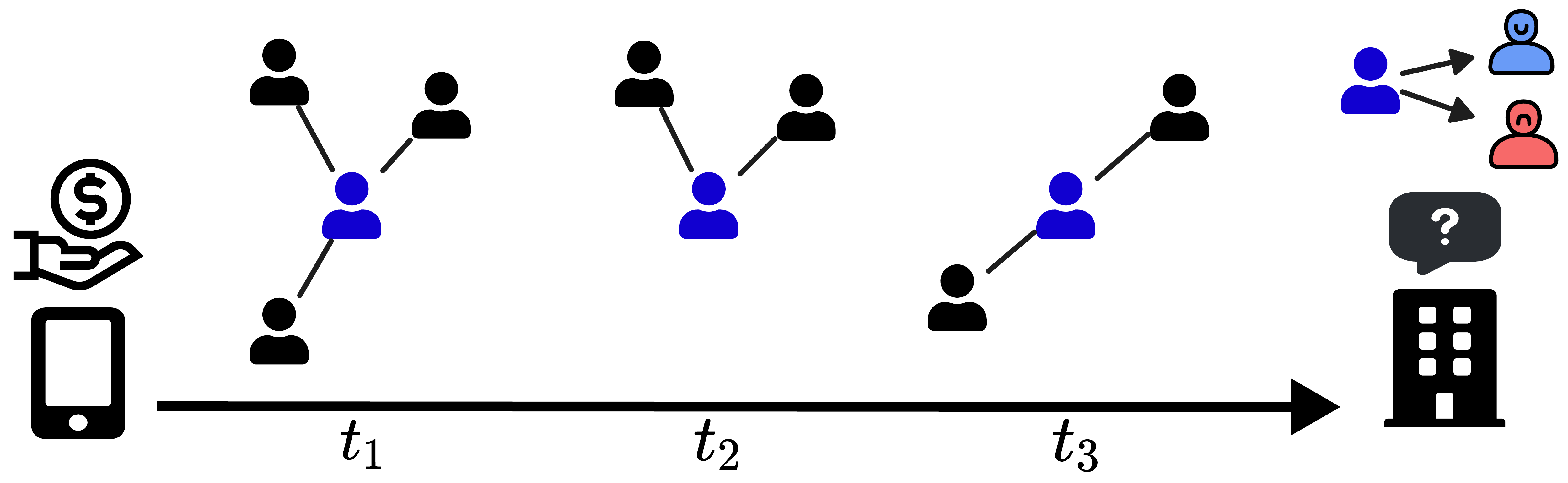}
    \caption{Visualization of temporally evolving edges in financial networks}
    \label{fig:fig}
\end{figure}

Graph analysis, and more specifically Graph Anomaly Detection (GAD), plays a crucial role in identifying irregular patterns or connections within these networks \cite{akoglu2015graph}. The continuous evolution of technology further amplifies the need for advanced research in financial fraud detection. Effective fraud detection systems are essential not only for mitigating financial losses but also for preserving the foundational trust that sustains the financial system.

Consider a scenario where a customer applies for a loan, and over time, the guarantor information linked to this customer changes. This situation, representative of the dynamic nature of financial networks, can be visualized as evolving connections in Figure \ref{fig:fig}. In such dynamic environments, it is imperative for financial institutions to rapidly and accurately distinguish between fraudulent and legitimate activities. Our paper addresses this need by exploring the application of Temporal Graph Network (TGN) \cite{rossi2020temporal} in enhancing the efficacy of GAD in financial networks.

\begin{figure*}[htp]  
    \centering
    \includegraphics[width=\textwidth, height=0.8\textheight, keepaspectratio]{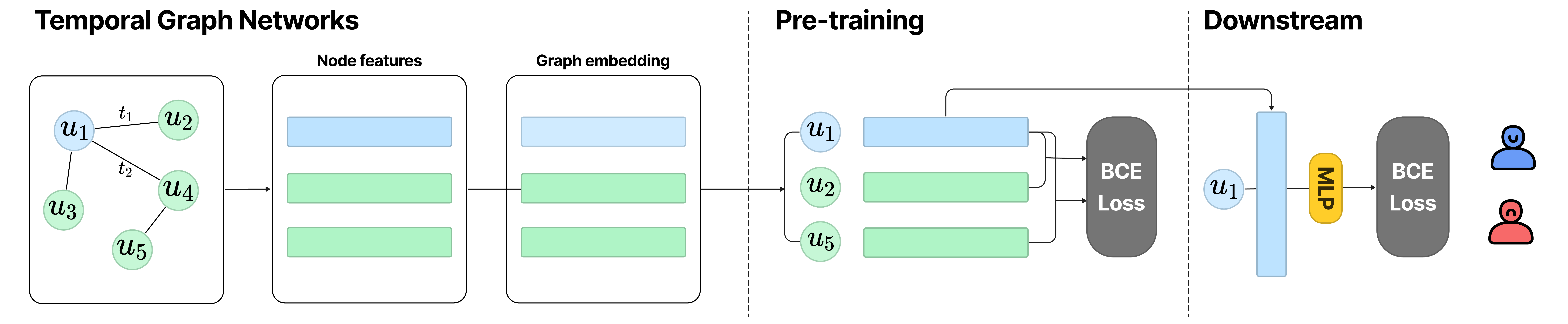}
    \caption{TGN architecture for graph anomaly detection in finance.}
    \label{fig:model}
\end{figure*}



TGN has emerged as a promising model capable of capturing these dynamic changes in nodes and edges effectively. TGN can learn from a graph that evolves over time, making it potentially suitable for applications like real-time fraud detection in financial transactions. However, the application of TGN for GAD in the finance domain is still not well-established.
Addressing this gap, our research presents a comprehensive framework for utilizing TGN in the financial domain, specifically for anomaly detection. We experimented with various models within this framework, comparing their performance against traditional static GNN models and hypergraph neural network baselines. Our findings reveal that TGN, with its ability to learn from dynamic edges, shows remarkable performance, demonstrating its potential as a powerful tool in financial fraud detection. This research not only advances the field of GAD in finance but also opens new avenues for the application of dynamic graph models in real-world scenarios.

\section{Preliminaries}
\subsection{Problem Definition}
A graph anomaly detection can be described as node classification task. From the financial network, we define a user node set $U = \{u_1, u_2, ... u_{|U|}\}$. The core challenge lies in discerning whether a node $u$ is `normal' user or `fraudster' user. 
This is represented by binary classification using node label $y_{u}$, where $y_{u}=0$ for normal users and $y_{u}=1$ for fraudster users. 
Ultimately, the goal is to accurately predict the state of $u$, enabling effective detection of anomalies of nodes.

\subsection{Continuous Time Dynamic Graph}
We construct a dynamic graph with user interactions, changing its structure over continuous time. We define our continuous-time dynamic graph as $\mathcal{G}(T)=(U, \mathcal{E}_T)$ where $\mathcal{E}_T$ denotes the temporal set of edges. Each edge in $\mathcal{E}_T$ is characterized by a tuple $e=(u_s,u_d,t)$, consisting of a source node $n_s$, an destination node $n_d$, and a timestamp $t$.

\section{Methodology}

\subsection{Temporal Graph Networks}
In our discussion, nodes are indexed by $i$, and neighboring nodes are indexed by $j$. The neighborhood set of node $i$, denoted as $n_i^k(t)$, refers to the $k$-hop temporal neighbors connected up to time $t$. We follow notations used in \cite{rossi2020temporal}.

We learn node embeddings from our dynamic graph, which are later used
for a downstream task, node classification. We build our model within the TGN framework excluding the memory module. This was due to the very large size of the data used in our experiments, leading to out-of-memory errors. In graph embedding module, temporal embeddings for a dynamic graph are generated, specifically creating embeddings for each node at time step $t$. We build various functions $f$ to effectively learn the connectivity between nodes. A node embedding can be represented as:
\begin{equation}
    \mathbf{z}_i(t)=\sum_{j \in n_i^k(t)} f\left(\mathbf{v}_i(t), \mathbf{v}_j(t), \mathbf{e}_{i j}\right)
\end{equation}

where $\mathbf{v}_i(t)$ and $\mathbf{v}_j(t)$ represent the node features, and $\mathbf{e}_{ij}$ is the edge feature. 

\subsubsection{\textit{Temporal Graph Attention} (attn)}
With function $f$ of attn, a node embedding $\mathbf{h}$ is initialized with the node's features as  $\mathbf{h}_i^{(0)}(t)=\mathbf{v}_i(t)$. Then, it learns the relationships between target nodes and neighboring nodes based on attention as it passes through $l$ layers.
\begin{equation}
    \mathbf{h}_i^{(l)}(t)=\text{MLP}^{(l)}\left(\mathbf{h}_i^{(l-1)}(t) \| \tilde{\mathbf{h}}_i^{(l)}(t)\right)
\end{equation}
\begin{equation}
\resizebox{\columnwidth}{!}{$
    \tilde{\mathbf{h}}_i^{(l)}(t)=\text{ MultiHeadAttention }{ }^{(l)}\left(\mathbf{Q}^{(l)}(t), \mathbf{K}^{(l)}(t), \mathbf{V}^{(l)}(t)\right)
$}
\end{equation}

where the the query $\mathbf{Q}$, key $\mathbf{K}$, and value $\mathbf{V}$ are defined as in TGAT \cite{velivckovic2017graph}.

\subsubsection{\textit{Temporal Graph Sum} (sum)}
With function $f$ of sum, a simpler aggregation over the graph is conducted. That is,


\begin{equation} 
\mathbf{h}_i^{(l)}(t) = \mathbf{W}_2^{(l)}\left(\mathbf{h}_i^{(l-1)}(t) \| \tilde{\mathbf{h}}_i^{(l)}(t)\right)
\end{equation}


\begin{equation} 
\resizebox{\columnwidth}{!}{$
\tilde{\mathbf{h}}_i^{(l)}(t) = \text{ReLU}\left(\sum_{j \in n_i([0, t])} \mathbf{W}_1^{(l)}\left(\mathbf{h}_j^{(l-1)}(t)\left\|\mathbf{e}_{i j}\right\| \boldsymbol{\phi}\left(t-t_j\right)\right)\right)
$}
\end{equation}

where $\boldsymbol{\phi}$ is a time encoding \cite{velivckovic2017graph}.

In addition to the two graph embedding methods proposed in \cite{rossi2020temporal}, we further experimented with two additional methods to explore the effectiveness of various graph embedding techniques.

\subsubsection{\textit{Temporal Graph Mean} (mean)}
Similar to sum, we can take mean of neighbor embeddings when aggregating information. 


\begin{equation} 
\resizebox{\columnwidth}{!}{$
\tilde{\mathbf{h}}_i^{(l)}(t) = \text{ReLU}\left(\frac{1}{|n_i([0, t])|}\sum_{j \in n_i([0, t])} \mathbf{W}_1^{(l)}\left(\mathbf{h}_j^{(l-1)}(t)\left\|\mathbf{e}_{i j}\right\| \boldsymbol{\phi}\left(t-t_j\right)\right)\right)
$}
\end{equation}

where $|n_i([0, t])|$ is the number of neightbor of $i$.

\subsubsection{\textit{Temporal Graph Convolution} (conv)}
Alternatively, we can distill contribution of neighbor embeddings by introduting a linear layer.


\begin{equation} 
\resizebox{\columnwidth}{!}{$
\tilde{\mathbf{h}}_i^{(l)}(t) = \text{ReLU}\left(\mathbf{W}_2^{(l)} \sum_{j \in n_i([0, t])} \mathbf{W}_1^{(l)}\left(\mathbf{h}_j^{(l-1)}(t)\left\|\mathbf{e}_{i j}\right\| \boldsymbol{\phi}\left(t-t_j\right)\right)\right)
$}
\end{equation}

For every graph embedding modules, the final node embedding is obtained from the last layer: $\mathbf{z}_i(t)=\mathbf{h}_i^{(L)}(t)$.

\subsection{Pre-training and Downstream task}
To conduct graph anomaly detection, we initially learn node embeddings through pre-training with the edge prediction task using TGN as an encoder. Subsequently, we train a separate MLP as a decoder for node classification as a downstream task by inputting the embeddings obtained from the encoder.

\subsubsection{Pre-training}
In this phase, we focus on a task that involves predicting connections between nodes. We posit that if a node $u$ is connected to node $i$ at time $t$, this scenario constitutes a positive pair. Conversely, a negative pair is formed by randomly sampling a node from the entire set of nodes. We denote such a negatively sampled node as $k$. To refine the model's ability to differentiate between these pairs, we utilize the Binary Cross-Entropy Loss (BCELoss):
The objective is to increase the probability of an edge existing in positive pairs while decreasing it for negative pairs. Mathematically, this can be represented as follows:

\begin{equation}
\resizebox{\columnwidth}{!}{$
\mathbf{L}_{1} = -\sum_{(u, i, k, t) \in D} [\log(z_u(t)^Tz_i(t)) + \log(1 - z_u(t)^Tz_k(t))]
$}
\end{equation}

where \(D\) denotes the edge set, which is derived from \(\mathcal{E}_T\) with additional negative sampling, \(u\) is the source node, \(i\) is positive destination node, \(k\) is negative destination node, and \(t\) represents time. 


\subsubsection{Downstream task}
In downstream phase, the objective shifts to predicting the class of each node. 
The predicted class for node $u$ is expressed as:

\begin{equation}
    y_u^{\text{pred}} = \text{MLP}(\mathbf{z}_u)
\end{equation}

where $\mathbf{z}_u$ is the node embedding at its final time. For the loss function in this phase, we employed the BCE loss:


\begin{equation}
\mathbf{L}_{2} = -\sum_{|U|} [y_u^{true} \log(y_u^{\text{pred}}) + (1 - y_u^{true}) \log(1 - y_u^{\text{pred}})]
\end{equation}

 where $y_u^{true}$ represents the true class of the nodes $u$. 

\section{Experiment}
\subsection{Experimental Settings}
\subsubsection{Dataset}

In our study, we used the DGraph dataset \cite{huang2022dgraph}, designed for fraud detection in real-world financial scenarios. Provided by Finvolution Group, a fintech platform, DGraph is built on interactions where the users fill in emergency contact of other users. It consists of 3,700,550 user nodes and 4,300,999 directed edges, making it a challenging GAD dataset with the extreme imbalanced classes.


\subsubsection{Baseline}
In our research, we employed a suite of baseline methods that are renowned in the graph neural network (GNN) domain, specifically chosen for their capability to handle the intricate structures and features of the DGraph dataset. Our baselines comprised MLP for its foundational neural network approach; Graph Convolutional Network (GCN) \cite{kipf2016semi}, which excels in aggregating neighbor information within graph structures; GraphSAGE \cite{hamilton2017inductive}, known for generating node embeddings by sampling and aggregating from a node's local neighborhood.

In addition to the aforementioned models, our suite of baseline methods encompasses the hypergraph neural networks. A hypergraph, defined by nodes and hyperedges, represents each hyperedge as a set of nodes, allowing it to capture higher-order relationships rather than pairwise relationships. We construct hyperedges by grouping destination nodes from each source node. In this setting, we implemented four distinct hypergraph neural networks. 
Hypergraph Neural Networks (HGNN) \cite{feng2019hypergraph} applies graph convolutions by transforming a given hypergraph into an ordinary graph. 
Hypergraph Networks with Hyperedge Neurons (HNHN) \cite{dong2020hnhn} first updates hyperedge embeddings using their including nodes and then updates node embeddings using their incident hyperedges.
Hypergraph Convolution and Hypergraph Attention (HCHA) \cite{bai2021hypergraph} utilizes parameters for the weights to the incident node or hyperedge embeddings during aggregation.
AllSetTransformer \cite{chien2021you} comprises two multiset functions with SetTransformer\cite{lee2019set} for aggregating node or hyperedge embeddings.


\subsubsection{Hyperparameter}
We conducted experiments with TGN models using the following configurations: a node embedding dimension of 128, a batch size of 200, and a learning rate of 0.0001 for pre-training. For the downstream task, we used a batch size of 100, and a learning rate of 0.0003.
For the graph baseline models, we consistently used a batch size of 1024. 
For hypergraph baselines, we performed a grid search within the following ranges: learning rate in 0.01, 0.02, 0.03, 0.04, 0.05, 0.06, 0.07, 0.08, 0.09, 0.1, 0.2, 0.3, 0.4, 0.5, and 0.6, hidden dimension in 16, 32, 64, and 128, number of layer in 1 and 2. 
All models were trained with 10 epochs.




\subsubsection{Metrics and Evaluation}
Our research employed one primary metric for model evaluation: Area Under the Curve (AUC). AUC is a widely recognized measure for the capability of a model to distinguish between different classes, in our case, between normal and fraudsters users. 
This metric was chosen for its robustness and widespread acceptance in evaluating the performance of classification models.

\subsection{Results and Discussion}
\begin{table}[h]
\centering
\renewcommand{\arraystretch}{1.2}
\begin{tabular}{|l||c|c|}
\hline

\textbf{Model} & \textbf{Valid AUC} & \textbf{Test AUC} \\ 
\hline
MLP            & 0.6507             & 0.6465            \\ 
GCN            & 0.6087             & 0.6135            \\ 
GraphSAGE      & 0.6344             & 0.6344            \\ 
\hline
AllSetTransformer & \underline{0.6779} & 0.6829          \\ 
HNHN           & 0.6775             & \underline{0.6845} \\ 
HGNN           & 0.6429             & 0.6450            \\ 
HCHA           & 0.6289             & 0.6318            \\ 
\hline
TGN(Sum)       & 0.7517             & 0.7640            \\ 
TGN(Mean)      & 0.7569             & \textbf{0.7747}   \\ 
TGN(GCN)       & \textbf{0.7581}    & 0.7716            \\ 
TGN(GAT)       & 0.7569             & 0.7691            \\ 
\hline
\textbf{\textit{improv.}} & 11.83\%       & 13.18\%           \\ 
\hline
\end{tabular}
\caption{AUC comparison between the models: the best-performing TGN model is highlighted in bold, and the top baseline model is underlined.}
\label{table:baseline}
\end{table}


As shown in Table \ref{table:baseline}, TGNs exhibit significantly better performance, confirming their effectiveness in aggregating information from a node's neighbor history. This enhanced ability contributes to superior sensitivity and specificity, particularly in identifying patterns indicative of anomalous behavior, compared to baseline models. 

Furthermore, we investigate various graph embedding modules within the TGN framework. While variants exist within TGN, the results consistently affirm the model's ability to achieve remarkable performance in the anomaly detection task. This contributes to a deeper understanding of the factors influencing TGNs' effectiveness in handling the complexity of emergency contact interactions.

Delving deeper into the baseline comparisons, it's noteworthy that MLPs exhibited better performance than static graph models like GCN and GraphSAGE. This outcome suggests that approaches which view interactions in an independent and static manner may not be as effective. The dynamic and interconnected nature of financial transactions poses a challenge to these static models, highlighting the need for more adaptive and dynamic approaches.

Additionally, the hypergraph baselines have demonstrated remarkable performance, offering a key insight: viewing interactions dependently is crucial in anomaly detection. This finding emphasizes the importance of considering the complex interdependencies inherent in financial transactions, which hypergraphs capture more effectively than traditional graph models.

\section{Conclusion}

In conclusion, the experimental results solidify the position of TGNs as a robust and promising approach for financial anomaly detection. The superior performance over baseline models, coupled with the adaptability demonstrated in capturing dynamic patterns, positions TGNs as a valuable asset in the arsenal against financial fraud.  These findings contribute not only to the academic discourse on anomaly detection but also hold practical implications for enhancing fraud detection measures in fintech applications.


Interestingly, the hypergraph baselines also exhibited commendable performance, suggesting its potential utility in anomaly detection tasks. Therefore, as a future work, we propose experimenting with a hybrid model that combines the strengths of TGN and hypergraph. This combination could potentially harness the temporal sensitivity of TGN and the intricate node relationship mapping of hypergraphs, offering a more holistic approach to anomaly detection.

\section{Acknowledgments}
This work was supported by Institute of Information \& communications Technology Planning \& Evaluation (IITP) grant funded by the Korea government(MSIT) (RS-2022-00143911, AI Excellence Global Innovative Leader Education Program)

\bibliography{aaai24}

\end{document}